\documentclass[preprint,showpacs,prX]{revtex4}

\input epsf
\usepackage{graphicx}
\usepackage{epsfig}
\usepackage{mathrsfs}
\usepackage{amsfonts}
\usepackage{mathrsfs}
\usepackage{natbib}
\usepackage{epstopdf}
\usepackage{subfigure}

\begin{document}

\title{Mechanics of Tunable Helices and Geometric Frustration in Biomimetic Seashells}

\author{Qiaohang Guo*$^{1,2,3}$, Zi Chen*$^{3}$, Wei Li$^{1,4}$, Pinqiang Dai$^{1,4}$, Kun Ren$^{4}$, Junjie Lin$^{4}$, Larry A. Taber$^{3}$ and Wenzhe Chen$^{1,4}$}
\affiliation{$^1$College of Materials Science and Engineering, Fuzhou University, Fujian 350108 China\\
$^2$Department of Mathematics and Physics, Fujian University of Technology, Fujian 350108 China\\
$^3$Department of Biomedical Engineering, Washington University in St. Louis, MO 63130\\
$^4$College of Materials Science and Engineering, Fujian University of Technology, Fujian 350108 China\\
* Qiaohang Guo and Zi Chen contributed equally to this work.\\
Electronic mail: guoqh@fjut.edu.cn; chen.z@seas.wustl.edu.
}

\date{\today}

\begin{abstract}
Helical structures are ubiquitous in nature and engineering, ranging from DNA molecules to plant tendrils, from sea snail shells to nanoribbons. While the helical shapes in natural and engineered systems often exhibit nearly uniform radius and pitch, helical shell structures with changing radius and pitch, such as seashells and some plant tendrils, adds to the variety of this family of aesthetic beauty. Here we develop a comprehensive theoretical framework for tunable helical morphologies, and report the first biomimetic seashell-like structure resulting from mechanics of geometric frustration. In previous studies, the total potential energy is everywhere minimized when the system achieves equilibrium. In this work, however, the local energy minimization cannot be realized because of the geometric incompatibility, and hence the whole system deforms into a shape with a global energy minimum whereby the energy in each segment may not necessarily be locally optimized. This novel approach can be applied to develop materials and devices of tunable geometries with a range of applications in nano/biotechnology.
\end{abstract}


\maketitle

\newcommand{\dx}{\mbox{${\bf d}_x$}}
\newcommand{\dy}{\mbox{${\bf d}_y$}}
\newcommand{\dz}{\mbox{${\bf d}_z$}}
\newcommand{\rx}{\mbox{${\bf r}_1$}}
\newcommand{\ry}{\mbox{${\bf r}_2$}}
\newcommand{\bN}{\mbox{${\bf N}$}}
\newcommand{\bP}{\mbox{${\bf P}$}}
\newcommand{\Ex}{\mbox{${\bf E}_x$}}
\newcommand{\Ey}{\mbox{${\bf E}_y$}}
\newcommand{\Ez}{\mbox{${\bf E}_z$}}
\newcommand{\dux}{\mbox{${\bf u}_1$}}
\newcommand{\duy}{\mbox{${\bf u}_2$}}
\newcommand{\drx}{\mbox{${\bf r}_1$}}
\newcommand{\dry}{\mbox{${\bf r}_2$}}
\newcommand{\strain}{\mbox{\boldmath $\gamma$}}
\newcommand{\andsp}{\mbox{$\quad\textrm{and}\quad$}}
\newcommand{\pfrac}[2]{\frac{\partial #1}{\partial #2}}
\newcommand{\Dfrac}[2]{\frac{\mathrm{d} #1}{\mathrm{d} #2}}

Helical structures are basic building blocks in biological and engineering systems, such as DNA\cite{Biton_2007}, plant tendrils \cite{Goriely_PRL1999, Goriely_PRL2006, Gerbode_Science2012, Wang_SR2013}, seashells\cite{Rice_Paleobiol_1998, Moulton_JTB_2012}, curly hair\cite{Bertails_2006}, cholesterol \cite{Chung_PNAS1993}, and nanoribbons \cite{kong_nano2003,Sawa_PNAS_2011}. Recent work has shown that helical ribbon morphology can be controlled by the mechanical balance between surface stresses or internal residual strains and the induced elastic stretching and bending \cite{Chen_APL2011, Armon_Science2011, Chen_ProcRSocA2012, Huang_SoftMatter2012, Gerbode_Science2012}. Importantly, the generation of helical morphology requires both mechanical anisotropy, such as anisotropy in surface stress \cite{Twisting_Wang2008APL}, residual strain, elastic properties \cite{zhang_nano2005}), and geometric mis-orientation between the principal mechanical axes and geometric axes (length, width, thickness) of the structure.

Noticeably, many of the helical structures studied with theoretical and experimental approaches exhibit uniform radius and pitch \cite{zhang_nano2005, Chouaieb_PNAS2006, Twisting_Wang2008APL, Armon_Science2011, Chen_APL2011, Sawa_2013}. Helical morphologies of variable radius, pitch and width are less heavily investigated. As typical representatives of helical shapes with geometric parameters, seashells have long fascinated scientists with their aesthetic beauty that roots in their self-similar, spiraling shapes with left-right asymmetry. Moseley first modeled the geometry of the coiling molluscan shell as a logarithmic spiral \cite{Moseley_PTRSL_1838}. Afterwards, a number of theoretical models of molluscan shells have been developed on the geometric properties of shell growth and morphogenesis \cite{Lovtrup_BMB_1974, Okamoto_Pala_1988, Stone_Palebio_1995}. In recent years, some theoretical models have addressed the biological processes taking place at the growing edge and the growth kinematics of shell aperture\cite{Checa_Lethaia_1991, Rice_Paleobiol_1998, Hammer_Lethaia_2005, Tyszka_Paleobiol_2005, Urdy_JEZBMDE_2010, Moulton_JOE_2011, Moulton_JTB_2012}. While the morphogenesis of seashells are not the main focus of the current study, it is of interest, from an engineering point of view, to develop both theoretical and experimental strategies of designing shapes inspired by nature, e.g., structures that mimics the seashells. In this letter, we first present a theoretical framework for helical morphologies with tunable geometric parameters such as principal radii of curvature, width and helix angle, but without self-contact. Then we address the mechanics of geometric frustration due to self-avoidance restriction, and report the generation of three-dimensional helical morphologies where the energy minimization is achieved on a global scale but not locally. According to this principle, we designed a helical, seashell-like structure through tabletop experiments. This study can promote understanding of morphogenesis in biological systems  \cite{Armon_Science2011}, and inspire new design principles for novel materials and devices of tunable morphologies or structures that can change configurations in response to external stimuli \cite{Ryu_APL2012, Sawa_PNAS_2011, Sawa_2013}, with applications in nanofabrication \cite{Suo_APL1999, kong_nano2003, zhang_nano2005} and bio-inspired technology \cite{abbott_ijrr2009, Godinho_softmatter_2010}.

\begin{figure}[t]
\begin{center}
\includegraphics[width=3.5in]{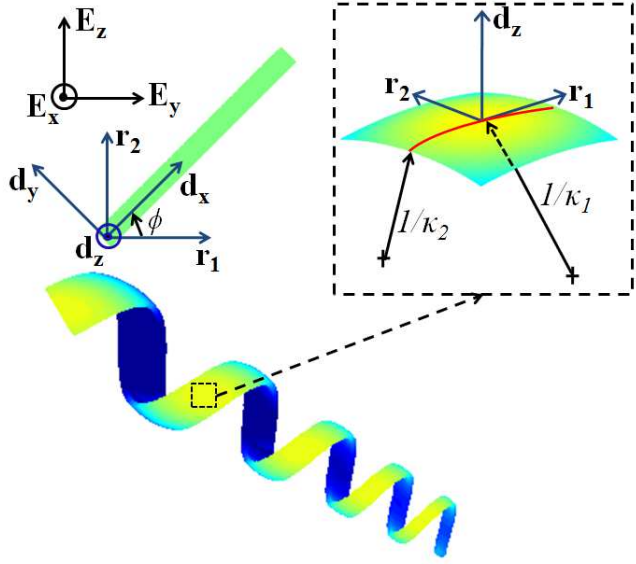}
\end{center}
\caption{Illustration of a helical ribbon with varying width and principal curvatures. The directors, $\dx$, $\dy$ and $\dz$, go along the length, widthwise and thickness direction of the ribbon respectively.  The bases ${\bf r}_1$ and ${\bf r}_2$ correspond to the principal axes of curvature. $\Ex$, $\Ey$ and $\Ez$ are the bases of the global Cartesian coordinate system.}
\label{fig1}
\end{figure}

In this work, the ribbon is considered as an elastic sheet with length $L$, width $w (s)\ll L$ (where $s$ is the arclength measured from the origin, sitting at the center of the ribbon's wider end), and thickness $H \ll w$.  The ribbon thus features rectangular cross-sections with changing width, and the principal geometric axes are along the length ($\dx$), width ($\dy$), and thickness ($\dz$) directions, which form an orthonormal triad, $\{\dx,\dy,\dz\}$, that convolutes with the bent and twisted ribbon in three-dimensional space.

Chen et al.'s recent works \cite{Chen_APL2011, Chen_ProcRSocA2012} have shown that the equilibrium configuration can be determined by locally minimizing the total potential energy, when the deformed ribbon has constant principal curvatures, and does not have any self-contact. Here, we first deal with the more general case of bi-axial bending with varying principal curvatures along the centerline, also without self-contact. In this case, the ribbon will have principle curvatures $\kappa_1 (s)$ and $\kappa_2 (s)$ along the directors ${\bf r}_1 = \cos\phi\dx - \sin\phi\dy$ and ${\bf r}_2 = \sin\phi\dx + \cos\phi\dy$ oriented at an angle $\phi$ relative to $\dx$ within the plane of the ribbon (see Fig. \ref{fig1}).  In the global cartesian coordinate system, the coordinates of a point $\textbf{P}(s)$ (parameterized by the arclength $s$) on the centerline can be obtained by integrating the following equations \cite{Chen_ProcRSocA2012}:
\begin{eqnarray}
\Dfrac{\bP}{s} &=&  \sin{\phi} \rx  + \cos{\phi} \ry,  \\
\Dfrac{\bN}{s} &=&  \kappa_1(s) \cos{\phi} \rx + \kappa_2(s) \sin{\phi} \ry \, \\
\Dfrac{\rx}{s} &=& - \bN \kappa_1(s) \cos \phi \, \\
\Dfrac{\ry}{s} &=& - \bN \kappa_2(s) \sin \phi \,.
\label{eq:kinematics}
\end{eqnarray}
where $\bN = \dz \equiv \dx\times\dy = \rx\times\ry$ is unit normal to the ribbon, together with the boundary conditions $\textbf{P}_0(s) = X_0(s) \Ex + Y_0(s) \Ey +  Z_0(s) \Ez\,, \bN(0) =  \Ez \,, \rx(0) = \cos\phi\Ex - \sin\phi\Ey \andsp \ry(0) = \sin\phi\Ex + \cos\phi\Ey$. It is worth noting that although analytic expressions can be obtained when $\kappa_1(s)$ and $\kappa_2(s)$ are both constant \cite{Chen_APL2011, Chen_ProcRSocA2012}, in the more general case of interest where $\kappa_1(s)$ and $\kappa_2(s)$ have an arclength dependance, the coordinates of the centerline can be solved numerically.

In the presence of bi-axial bending curvatures, the deformed ribbon exhibit strain components $\epsilon_{xx}$, $\epsilon_{yy}$, $\epsilon_{xy}$, and $\epsilon_{zz}$ which are considered uniform throughout the $\dy$ direction of the ribbon (when geometric nonlinearity is relatively week). More general consideration involving geometric nonlinearity can be done following the recent work of Chen et al.\cite{Chen_PRL2012}. Here, we choose not to include the nonlinear geometric effects for the clarity of statement about the procedure.  By superposition we obtain the strain tensor $\strain = \gamma_{ij}{\bf d}_i\otimes{\bf d}_j$ ($i,j \in \{x,y,z\}$) with components
\begin{eqnarray}
\gamma_{xx} &=& \epsilon_{xx} + z (\kappa_1 \cos^2 \phi + \kappa_2 \sin^2 \phi)+\gamma_{xx}^0(z) \nonumber\\
\gamma_{xy} &=& \epsilon_{xy} + z (\kappa_2 - \kappa_1) \sin \phi \cos \phi+\gamma_{xy}^0(z) \nonumber\\
\gamma_{yy} &=& \epsilon_{yy} + z (\kappa_1 \sin^2 \phi + \kappa_2 \cos^2 \phi)+\gamma_{yy}^0(z) \nonumber\\
\gamma_{zz} &=& \epsilon_{zz} + z k_3+\gamma_{zz}^0(z) \,.
\end{eqnarray}
Here, $z \in [-H/2,H/2]$ is the distance of any point in the ribbon away from the midplane, while $\gamma_{ij}^0(z)$ represents the residual strain component within the initially flat ribbon. On the top and bottom surfaces ($z = {\pm}H/2$), effective surface stresses ${\bf f}^{\pm}$ serve as the driving forces for spontaneous deformation. The potential energy per unit area in the ribbon is $\Pi = {\bf f}^-:\strain |_{z=-H/2} + {\bf f}^+:\strain |_{z=H/2} + \int_{-H/2}^{H/2}\frac{1}{2}\strain:{\bf C}:\strain\,dz\,,$, where ${\bf C}$ denotes the fourth-order elastic constant tensor.  The equilibrium configuration can be achieved by minimizing the potential energy $\Pi$: $\partial\Pi/\partial\chi = 0$, where $\chi$ represents any of the undetermined parameters \cite{Chen_APL2011}.

Next, we consider the more interesting scenario with potential self-contact \cite{Chouaieb_PNAS2006}. For example, for an intrinsically helical ribbon with a linearly varying width, $W = W_0 - \alpha s$ (or constant, i.e., $\alpha = 0$), where the minimal energy shape cannot be achieved due to the potential self-contact restriction, what will the possible equilibrium configuration be?
\begin{figure}[t]
\begin{center}
\includegraphics[width=2in]{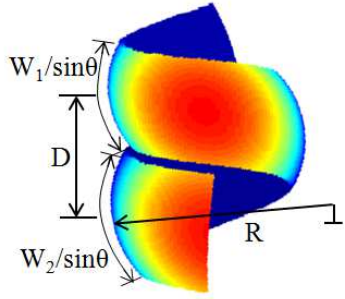}
\end{center}
\caption{Illustration of one helical turn of a non-self-overlapping ribbon with varying width, but constant principal curvatures.}
\label{fig1b}
\end{figure}
More specifically, given a virtually equilibrium configuration with natural principal curvatures in the forbidden regime (i.e., with self-penetration), what will the actual minimum energy configuration be? In a pioneering work, Chouaieb and co-workers \cite{Chouaieb_PNAS2006} studied the specific case for the self-contacting case of a uniform helical rod. Here, the geometry and mechanics involved are more complicated, since the ribbon can have a varying width, and non-constant principal curvatures. We first do local energy minimization to find the equilibrium shape without considering self-avoidance, i.e., when there is no energy penalty for doubly occupying the same space. For a segment of more than one complete turn, however,  self-contact becomes inevitable when the projected distance (along the helix axis direction) between adjacent turns are smaller than the pitch, i.e.,
\begin{equation}
R \left[\sin{(\frac{W_1}{2R \sin\theta})} + \sin{(\frac{W_2}{2R \sin\theta})}\right] \leq D,
\label{eq:selfcontact}
\end{equation}
where $D = 2\pi (\kappa_1 - \kappa_2) \sin\phi \cos\phi/(\kappa_1^2 \cos^2{\phi} + \kappa_2^2 \sin^2{\phi})$ is the pitch, $R = 1/[\kappa_1^2 \cos{(\phi + \theta)} + \kappa_2^2 \sin{(\phi + \theta)}]$ is the radius of curvature along the helix axis direction (see Fig.\ref{fig1b}), and $\theta = \arctan{[(\kappa_1 - \kappa_2)\cos\phi \sin\phi/(\kappa_1^2 \cos{\phi} + \kappa_2^2 \sin{\phi})]}$ is the angle between the helix axis and the longitudinal axis of the ribbon. When the radii of curvature along the ribbon are such that the self-contact happens everywhere, a tightly-coiled ribbon results. This is similar to, but more complicated than, the example of a stress-free helical rod that cannot achieve its equilibrium configuration due to self-avoidance constraint \cite{Chouaieb_PNAS2006}. The strain-energy function for a hyperelastic ribbon is $W_{\texttt{tot}} = \int W(\strain - \strain^*) d\textbf{r} = \int \frac{1}{2} (\strain - \strain^*)\cdot K (\strain - \strain^*) d\textbf{r}$, where $\strain^*$ are the strains in the unstressed reference configuration where there are no resultant moments everywhere, $\textbf{r}$ is the vector of a point in the ribbon, and $K$ is a 3-by-3 symmetric positive-definite matrix \cite{Chouaieb_PNAS2006}. The minimum-energy configuration can be found by first constructing the energy level sets and then tracing the tangential points between the energy level set and the in-accessible (forbidden) region due to self-penetration. In principle, there are two tangential points, indicating two possible solutions (local minima) with opposite handedness. In reality, however, the configuration with smaller energy will be the global equilibrium featuring a preferred handedness.

\begin{figure}[t]
\begin{center}
\includegraphics[width=6in]{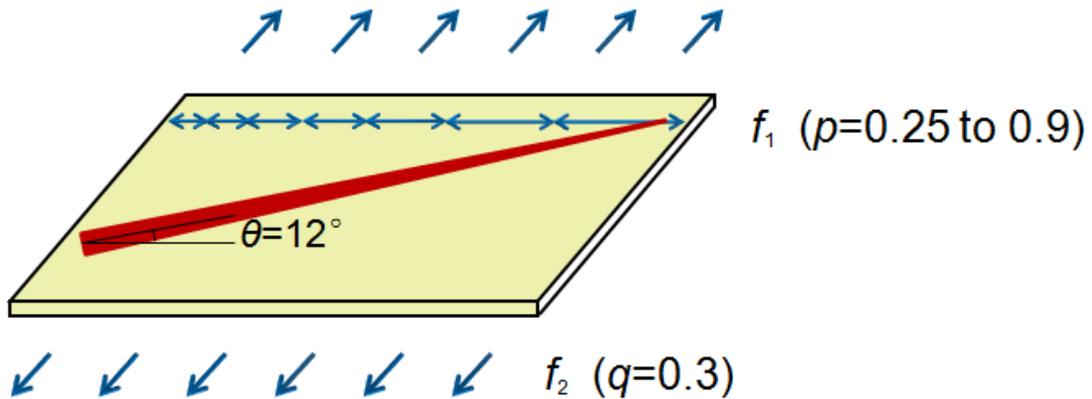}
\end{center}
\caption{Schematics for fabrication of a biomimetic seashell ribbon. A piece of latex rubber is bi-axially pre-stretched before bonded to an unstrained elastic adhesive sheet and subsequently cut into a triangular shape of length $360$mm with the width linearly varying between $13 $mm and $2$mm, with a mis-orientation angle $\phi = 12^{o}$. The bonded bilayer system, upon release, deformed into a tightly coiled seashell shape, as shown in Fig.\ref{fig4}.
}
\label{fig3}
\end{figure}

As a proof of concept, we designed table-top experiments to manufacture biomimetic seashells (\emph{Turritella}). 
\emph{Turritella} is a common kind of dextral seashell species, and the typical surface of the shell is shown in Fig. \ref{fig3}(a). The radius and width of the seashell are constantly changing, and the adjacent turns are in contact.

\begin{figure}[t]
\begin{center}
\includegraphics[width=3in]{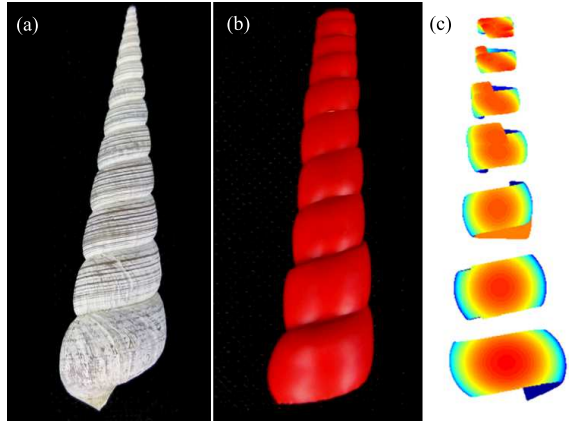}
\end{center}
\caption{Photographs of a real seashell and a biomimetic seashell ribbon. (a) Surface of a seashell where the spiral shell features constantly varying principal curvatures. (b) A piece of latex rubber (solid lines) is bi-axially pre-stretched before bonded to an unstrained elastic adhesive sheet and subsequently cut into  a triangular shape with the width linearly varying between $12 $ mm and $2 $ mm, with a mis-orientation angle $\phi$. The bonded bilayer system, upon release, deformed into a tightly coiled seashell shape. (c) Simulation results for each segment (without penalty for self-contact) of the mimetic-seashell ribbon showing the change of handedness as detailed in the text.
}
\label{fig4}
\end{figure}

Considering the feature of the seashell surface as shown in Fig. \ref{fig4}(a), we designed a table-top experiment so as to set up an energy hypothesis of bio-mimetic seashell structure. In our experiments, two sheets of latex rubber were pre-stretched and bonded to an elastic strip of thick, pressure-sensitive adhesive~\cite{Chen_APL2011}. One piece of thin latex rubber sheet (thickness $H_1$, and length $L$) was pre-stretched bi-axially and bonded to an unstrained triangular elastic strip of thicker, pressure-sensitive adhesive along a mis-orientation angle with respect to the principal axes (see the methods in \cite{Chen_APL2011, Chen_PRL2012}), such that the total thickness of the bonded strip is $H = H_1+H_2$. Here, the mis-orientation angle was $\phi = 12^{\texttt{o}}$, the pre-strain along the $\dy$ direction was 0.3, while the pre-strain along the changed from 0.25 to 0.90 in a piecewise manner, so that in the segments (in Fig. \ref{fig4}(c)) the pre-strain are 0.25, 0.3, 0.4, 0.5, 0.6, 0.7, and 0.9, respectively. The horizontal lengths of the segments after stretching are $50$mm, $50$mm, $40$mm, $40$mm, $30$mm, $20$mm, and $20$mm, respectively. 
Upon release, the bonded composite sheet deformed into seashell-like shapes (see Fig. \ref{fig4}(b)).

If the adjacent turns were allowed to overlap without additional energy cost, the configurations of each segment would have conformed to those shown in Fig. \ref{fig4}c. It is noteworthy that all the segments would come into self-contact if the condition in Eq. (\ref{eq:selfcontact}) is satisfied, and that the chirality would change from right-handed to left-handed. This change of chirality can be naturally interpreted using the recently developed elasticity model \cite{Chen_APL2011, Chen_ProcRSocA2012}, whereby the handedness is given by the sign of the helix angle, $\Phi = \arctan (\kappa_1 - \kappa_2) \sin \phi \cos \phi/(\kappa_1 \cos^2{\phi} + \kappa_2 \sin^2{\phi})$. When the first principal curvature is smaller than the second, the helical ribbon is left-handed (the bottom segment in Fig. \ref{fig4}(c)); while the first principal curvature becomes smaller than the second, the helical shape becomes right-handed (the top five segments in Fig. \ref{fig4}(c)); a ring-like shape results when the two principal curvatures are equal (the second last segment in Fig. \ref{fig4}(c)). But since self-penetration will lead to an infinite energy penalty, and hence not allowed, the overall geometric compatibility requires that the ribbon conform to a seashell-like shape as shown in Fig. \ref{fig4}(b), featuring a right-handed, tightly-coiled configuration, consistent with the theoretical prediction (the predominant handedness in the unstressed configuration prevails).

In sum, we develop a comprehensive theoretical framework for spontaneous helical ribbon structures with tunable geometric parameters and no self-contact by employing continuum elasticity, differential geometry, and stationary principles. Moreover, for ribbons that cannot access the locally stable shapes due to self-penetration, we show that a tightly coiled helical shape can result with a preferred handedness so that the total energy is globally minimized but not locally. Based on this principle, we designed a helical, seashell-like structure through simple tabletop experiments. This study represents a new paradigm for predicting and prescribing helical structures, and can inspire new design principles for novel materials and devices of tunable morphologies with broad applications in biological and engineering practises.

{\it Acknowledgements} -- The authors would like to thank Yushan Huang, Zhen Liu, Si Chen for their assistance in the experiments. This work has been in part supported by National Natural Science Foundation of China (Grant No.11102040), Projects of International Cooperation and Exchanges NSFC (Grant No.11201001044), Foundation of Fujian Educational Committee (Grant No.JA12238), the Sigma Xi Grants-in-Aid of Research (GIAR) program, American Academy of Mechanics Founder's Award from the Robert M. and Mary Haythornthwaite Foundation, and Society in Science, The Branco Weiss Fellowship, administered by ETH Z$\ddot{u}$rich (Z.C.).

\end{document}